\documentclass[a4paper,english,aps,floats,onecolumn,showpacs,nofootinbib]{revtex4}
\usepackage{pslatex}
\usepackage[T1]{fontenc}
\usepackage{graphicx}
\usepackage{epsfig}

\usepackage{calc}
\usepackage{ifthen}
\usepackage{subfigure}

\newcommand{\bea}{\begin{eqnarray}}
\newcommand{\eea}{\end{eqnarray}}

\newcommand{\nc}{\newcommand}
\nc{\renc}{\renewcommand}
\nc{\eqs}[2]{\mbox{Eqs.~(\ref{#1},\,\ref{#2})}}
\nc{\eq}[1]{\mbox{Eq.~(\ref{#1})}}
\nc{\figs}[2]{\mbox{Figs.~(\ref{#1},\,\ref{#2})}}
\nc{\fig}[1]{\mbox{Fig~.(\ref{#1})}}
\nc{\be}[1]{\begin{equation} \mbox{$\label{#1}$}}
\nc{\ee}{\vspace{0.1cm}\end{equation}}

\newcommand{\bean}{\begin{eqnarray*}}
\newcommand{\eean}{\end{eqnarray*}}

\def\GeV{{\rm \ GeV}}
\def\MeV{{\rm \ MeV}}

\def\TeV{{\rm \ TeV}}

\def\lae{\;^{<}_{\sim} \;} \def\gae{\; ^{>}_{\sim} \;}

\def\pb{\phi_{B}}

\def\pbh{\hat{\phi}_{B}}

\def\sp{\hat{s}}

\def\lhs{\lambda_{s}}

\nc{\npp}[3]{{\it  Nucl.\ Phys.\ }{{\bf #1} {(#2)} {#3}}}
\nc{\prdd}[3]{{\it  Phys.\ Rev.\ D\ }{{\bf #1} {(#2)} {#3}}}
\nc{\prll}[3]{{\it Phys.\ Rev.\ Lett.\ }{{\bf #1} {(#2)} {#3}}}
\nc{\pll}[3]{{\it  Phys.\ Lett.\ }{{\bf #1} {(#2)} {#3}}}

\begin{document}
\title{Simultaneous Generation of WIMP Miracle-like Densities of Baryons and Dark Matter
}
\author{John McDonald}
\email{j.mcdonald@lancaster.ac.uk}
\affiliation{Department of Physics, University of
Lancaster, Lancaster LA1 4YB, UK}

\begin{abstract}

      The observed density of dark matter is of the magnitude expected for a thermal relic weakly-interacting massive particle (WIMP). In addition, the observed baryon density is within an order of magnitude of the dark matter density. This suggests that the baryon density is physically related to a typical thermal relic WIMP dark matter density. We present a model which simultaneously generates thermal relic WIMP-like densities for both baryons and dark matter by modifying a large initial baryon asymmetry. Dark matter is due to O(100) GeV gauge singlet scalars produced in the annihilation of the O(TeV) coloured scalars which are responsible for the final thermal WIMP-like baryon asymmetry. The requirement of no baryon washout implies that there are two gauge singlet scalars. The low temperature transfer of the asymmetry to conventional baryons can be understood 
if the long-lived O(TeV) coloured scalars have large hypercharge, $|Y| > 4/3$. Production of such scalars at the LHC would be a clear signature of the model.

 \end{abstract}
\maketitle

\section{Introduction}

      The observed density of baryons and of dark matter are  within an order of magnitude of each other. If we discount simple coincidence as an explanation, there are broadly two approaches to understanding the baryon to dark matter ratio. One is the simultaneous production of baryons and dark matter, usually via decay of a particle to similar number of baryons and dark matter particles, as would be expected if there was a conserved charge carried by both \cite{direct0, direct1, direct2, direct3, direct4}. (This is closely related to models of asymmetric dark matter, which have been a focus of recent interest \cite{adm}.) In this case we expect $n_{DM} \sim n_{B}$ and so $m_{DM} \sim m_{n} = 1  \GeV$. (Models exist which break this simple relation \cite{jm1,vdirect1,vdirect2,vdirect3,vdirect4,vdirect5,vdirect6}.) The other is anthropic selection. An example of this is the case of axion dark matter, where superhorizon domains with different dark matter densities can be generated and $\rho_{DM} \sim \rho_{B}$ may then be anthropically favoured by the baryon density in galaxies \cite{lindeax}.

   However, these approaches generally neglect the other  notable coincidence of the dark matter density, its similarity to the thermal relic density of particles whose mass and interactions are characterized by the weak scale, the so-called "WIMP miracle". If the WIMP miracle is not a coincidence but instead an indication of the process responsible for dark matter, and if we discount anthropic selection, then the baryon asymmetry must also be related in some way to the WIMP miracle. 

   Thus there are three possibilities (excluding coincidence): (i) the WIMP miracle is the origin of the 
dark matter density and the baryon asymmetry is physically related to the WIMP miracle,  (ii) the WIMP miracle is the origin of the dark matter density and the baryon asymmetry is related to this by anthropic selection, and (iii) the WIMP miracle is not the explanation of dark matter. 

   The question of whether (i) is possible is therefore fundamentally important. If such a mechanism exists, it would be possible to understand both of the coincidences of the dark matter and baryon densities, why they are related to each other and to the WIMP miracle, in terms of particle physics. If not, it would tell us that either (ii) or (iii) is true i.e. either anthropic selection plays an essential role or the WIMP miracle is just a coincidence, not related to the origin of dark matter.

   In \cite{bm1} we proposed a model which could account for a thermal WIMP-like density of baryons by modifying a large initial baryon asymmetry via a weak-strength B-violating annihilation process, a process we call
baryomorphosis. The baryon asymmetry is initially locked in a density of particles which are decoupled from the thermal Standard Model (SM) background \cite{bm1}. These particles decay to  pairs of scalar particles $\phi_{B}$, $\hat{\phi}_{B}$ of mass O(1) TeV ('annihilons'). $\pb$ and $\pbh$ have opposite gauge charges but, importantly, not 
opposite baryon number. They annihilate to final state scalars via a B-violating, naturally weak-strength interaction. If the temperature at which the baryon number 
is transferred to annihilons is less than the freeze-out temperature of the B-violating interaction, a non-thermal but thermal WIMP-like relic density of $\pb$ and $\pbh$ will remain. These subsequently decay to conventional baryons, 
 leaving a baryon density which is naturally similar to a thermal relic WIMP dark matter density. 
  
 While the original model in \cite{bm1} demonstrated that it might be possible to understand why the baryon asymmetry is similar to a thermal relic WIMP density, it also highlighted some obstacles to be overcome in the construction of a natural model. Since the question we wish to answer is whether  
there exists a plausibly natural extension of the Standard Model (SM) which can account for a thermal WIMP-like baryon asymmetry, the naturalness of its construction is an important issue. 

    One issue is the danger of washout of the baryon asymmetry by the B-violating annihilation process. This excluded the possibility that $\pb$ and $\pbh$ could couple to the Higgs bilinear $H^{\dagger}H$, since this would induce a B-violating mixing of $\pb$ and $\pbh$ once the Higgs VEV was included, leading to baryon washout via scattering of $\pb$ and $\pbh$ from the thermal background \cite{bm1}. Therefore $\pb$ and $\pbh$ must only couple to scalars which do not have a VEV. In addition, a tree-level $\pb \pbh$ mixing term must be excluded and loop corrections should not generate a dangerous mixing between $\pb$ and $\pbh$. We will show how such dangerous terms may be excluded via a simple discrete symmetry.

   A second issue concerns the decay of $\pb$ and $\pbh$ to conventional 
baryons. This must occur after the $\pb$ $\pbh$ density annihilates to its final form, which occurs at the decay temperature of the initial asymmetry $T_{d}$. Since typically $T_{d} \lae 100 \GeV$ when the B-violating process is due to TeV scale particles, the lifetime of $\pb$ and $\pbh$ must be long, $\gae 10^{-10}$ s. We therefore need to understand why large renormalizable couplings between $\pb$ and $\pbh$ and conventional SM fermions, which would lead to rapid decay, are excluded.  

    In the original baryomorphosis model \cite{bm1}, the nature of the dark matter particle was not addressed; it was simply assumed to be a conventional WIMP. 
However, since we need to introduce new scalar particles to serve as the final state of the B-violating annihilation process and a new discrete symmetry to control B-washout, a natural possibility is that these new scalar particles could account for dark matter which is stabilized by a discrete symmetry. In this case the decay of the large initial baryon asymmetry can lead to a final baryon asymmetry and a dark matter density which are both similar to a typical thermal relic WIMP density.

     In this paper we will present a simple scalar extension of the SM which can account for thermal relic WIMP-like densities of both baryons and dark matter. The paper is organized as follows. In Section 2 we specify the model and its discrete symmetries. In Section 3 we discuss the modification of the large initial baryon asymmetry to a thermal WIMP-like baryon asymmetry and the production of a scalar dark matter density. In Section 4 we present the baryon asymmetry and dark matter density as a function of the masses and couplings of the scalars. In Section 5 we discuss the transfer of the baryon asymmetry to conventional baryons. In Section 6 we present our conclusions. In the Appendices we discuss the slowing of the annihilons and gauge singlet scalars by scattering from the thermal background and we provide the annihilation cross-section for the gauge singlet scalars.

\section{The Model}

       The model is based on a pair of scalars ('annihilons') $\pb$ and $\pbh$, with mass O(TeV) and with opposite gauge charge. (As we will discuss, although it is possible to construct a model with gauge singlet annihilons, the annihilons must carry either a global or gauge charge, strongly suggesting that they have SM gauge charges.)  To be specific, we will focus on the case where the annihilons $\pb$ and $\pbh$ are colour triplets, transforming as $({\bf 3}, {\bf 1})$ and $({\bf \overline{3}}, {\bf 1})$ under $SU(3)_{c} \times SU(2)_{L}$; other charge assignments will have a  similar cosmology.

     We first consider the symmetries that are required to evade $\pb$ $\pbh$ mixing that could lead to baryon washout. In \cite{bm1} the B-violating interaction allowing the annihilation of the annihilons was assumed to be of the form 
\be{e1} {\cal L}_{\phi_{B}\pbh \; ann} = \lambda_{B} \phi_{B} \pbh \hat{H}^{\dagger}\hat{H}  \;\;\; + \;h. \; c.  ~.\ee
Here $\hat{H}$ is a scalar which develops no expectation value. However, such an interaction means that a mixing term of the form $\Delta m \pb \pbh$
cannot be excluded by any symmetry.
There is also no symmetry which can exclude an interaction with the Higgs of the form $\lambda \pb \pbh H^{\dagger}H$, generating a mixing term with $\Delta m^2 = \lambda_{B} <H>^2$. Such mixing terms will generally lead to washout of the baryon asymmetry for natural values of the couplings, for example via scattering from the thermal background via gauge boson exchange, which imposes the constraint $\Delta m \lae 0.1 \GeV$ \cite{bm1}. This requires that $\lambda_{B} \lae 10^{-7}$. Such couplings are not consistent with a thermal WIMP-like baryon asymmetry from annihilation of TeV-scale particles via \eq{e1}, which requires $\lambda_{B} \sim 0.1$.

   Moreover, the interaction \eq{e1} will lead to a quadratic divergent term of the form $\Delta m \pb \pbh$.  Quadratic divergent contributions are not necessarily a problem. The situation is similar to the case of the Higgs boson mass in the SM. This is treated as a phenomenological input and the quadratic divergence is absorbed into the physical mass by renormalization. The same can be true for $\Delta m$.  However, such a solution is not acceptable if the theory is considered a low energy effective theory with a physical cutoff $\Lambda \gae 1 \TeV$. In this case the quadratic divergence would be considered a real contribution to $\Delta m$, requiring $\lambda_{B} \lae 10^{-6}$.

   These problems can be solved if the product $\pb \pbh$ transforms under a discrete $Z_{2}$ symmetry such that $\pb \pbh \rightarrow - \pb \pbh$. This can be achieved by a introducing a discrete symmetry $Z_{A}$ and real scalars $s$ and $\sp$, where $Z_{A}$ is defined by  
\be{e2} \pb \rightarrow \pb  \; ; \;\;\; \pbh \rightarrow -\pbh   \; ; \;\;\;   s \rightarrow s  \; ; \;\;\;  \sp \rightarrow -\sp  ~,\ee 
with all SM fields invariant under $Z_{A}$.    
$Z_{A}$ excludes terms of the form $\pb \pbh$ and $\pb \pbh H^{\dagger} H$ but allows the interaction term
\be{e3} {\cal L}_{\phi_{B}\pbh \; ann} = \lambda_{B} \phi_{B} \pbh s \sp     \;\;\; + \;h. \; c.  ~.\ee
It also excludes the dangerous interactions $\pb \pbh s s$ and $\pb \pbh \sp \sp$, which would generate quadratic divergent $\pb \pbh$ mixing terms.

    Note that if $\pb$ and $\pbh$ were gauge singlets, the B-violating mass term $\pb \pb$ would not be excluded, nor would the term $\pb \pb s s$ which leads to a quadratic divergent mixing. 
To exclude these, $\pb$ and $\pbh$ must also be oppositely charged with respect to either a global or a gauge symmetry. It is therefore natural to assume that $\pb$ and $\pbh$ carry SM gauge charges.

    The SM requires a dark matter candidate. In the context of the present model the simplest possibility is to consider one or both of $s$ and $\sp$ to be dark matter particles. We will consider the simplest case where 
$s$ and $\sp$ are real gauge singlet scalars. (For discussions of gauge singlet scalar dark matter see \cite{gsdm1,gsdm2,gsdm3,gsdm4,gsdm5}.) Models based on complex singlets or inert doublets \cite{id} could also be constructed. To ensure that the gauge singlet scalars are stable dark matter particles, we introduce an additional discrete symmetry $Z_{S}$, under which $s$ and $\sp$ are odd and all other particles are even.  We then expect couplings to the SM of the form 
\be{e4} \frac{\lambda_{s}}{2} s s H^{\dagger}H + \frac{\lambda_{\sp}}{2} \sp \sp H^{\dagger} H      ~.\ee 
These couplings will allow the $s$ and $\sp$ densities resulting from annihilation of the large initial $\pb$ and $\pbh$ density to annihilate down to a thermal relic WIMP-like densities.
For simplicity, we will consider $s$ and $\sp$ to have the same mass $m_{s}$ and the same Higgs coupling $\lambda_{s} = \lambda_{\sp}$. In this case there are two dark matter scalars, both with the same density.

\section{Baryon and Dark Matter Densities}

   We first give an overview of the process. As in \cite{bm1}, we will consider the decay of a large baryon asymmetry, initially locked in a density of thermally decoupled heavy particles, to a baryon asymmetry in relativistic annihilons $\pb$ and $\pbh$ at a low temperature $T_{d}$. 
(For simplicity we assume the annihilons have equal mass, $m_{\pb} = m_{\pbh}$.) 
$T_{d}$ should be less than the freeze-out temperature $T_{\pb}$ of the non-relativistic $\pb \pbh$ annihilation process due to \eq{e3}, in order that B-violation due to \eq{e3} does not come into thermal equilibrium. As discussed in Appendix A, the annihilons rapidly lose energy by scattering from the thermal background, becoming non-relativistic before there is any significant change in temperature from $T_{d}$. Once non-relativistic, they will annihilate via \eq{e2} to a residual annihilon asymmetry and to equal densities of $s$ and $\sp$. 

      The gauge singlets $s$ and $\sp$ from $\pb \pbh$ annihilation are initially relativistic. They become rapidly non-relativistic via t-channel Higgs exchange scattering with thermal background particles provided that $T_{d} \gae 0.4 \GeV$, in which case relativistic c-quarks form part of the thermal background (Appendix A). If $T_{d} < T_{s}$, where $T_{s}$ is the freeze-out temperature of   
the $s$ and $\sp$ scalar annihilation process from \eq{e4}, the $s$ and $\sp$ densities will annihilate down to non-thermal but thermal WIMP-like 
relic densities\footnote{If $T_{\pb} > T_{d} > T_{s}$, which can occur of $m_{s}$ is sufficiently light, $s$ and $\sp$ will have purely thermal relic densities. However, as we will discuss, this requires that the $s$ and $\sp$ masses are close to the Higgs pole.}

    Thus both the baryon asymmetry and dark matter densities will be fixed by non-relativistic annihilation processes at $T_{d}$. Since the annihilation processes from \eq{e3} and \eq{e4} are 
broadly of weak interaction strength when $m_{\pb}$ and $m_{s}$ are in the range O(100) GeV - O(1) TeV and when $\lambda_{B}$  and $\lambda_{s}$ are O(0.1), which are natural assumptions for an extension of the SM, the resulting non-thermal baryon asymmetry and dark matter density will be naturally similar to each other and to a thermal relic WIMP density as long as $T_{d}$ is not very small compared  to the freeze-out temperatures $T_{\pb}$ and $T_{s}$ \cite{bm1}.

\subsection{Baryon asymmetry from $\pb$ $\pbh$ annihilation} 

    The non-relativistic annihilation cross-section times relative velocity for the process $\pb \pbh \rightarrow s \sp$ from \eq{e3} is 
 \be{e5}    <\sigma v>_{\pb} = \frac{\lambda_{B}^2}{32 \pi m_{\pb}^2}
 \left( 1- \frac{m_{s}^{2}}{m_{\pb}^{2}} \right)^{1/2}    ~.\ee
The freeze-out number density at $T_{d}$ is then 
\be{e6}  n_{\pb}(T_{d}) \approx \frac{H(T_{d})}{<\sigma v>_{\pb}}    ~.\ee
(The $\pbh$ number density is the same.) 
If, as discussed later, $\pb$ decays to baryon number $B(\pb)$ and $\pbh$ to $B(\pbh)$, the baryon asymmetry to dark matter ratio at present, $r_{BDM} \equiv \Omega_{B}/\Omega_{DM}$, is given by 
\be{e6a}  r_{BDM} = 3 (B(\pb) + B(\pbh)) \frac{m_{n}}{\Omega_{DM}} \frac{g(T_{\gamma})}{g(T_{d})^{1/2}} \left( \frac{4 \pi^3}{45 M_{Pl}^2} \right)^{1/2} \frac{T_{\gamma}^{3}}{\rho_{c}}  \frac{1}{T_{d}} \frac{1}{\left< \sigma v\right>_{\pb}}   ~.\ee
Here $g(T)$ is the number of relativistic degrees of freedom in 
equilibrium, $m_{n}$ is the nucleon mass, $\rho_{c}$ the critical density, $T_{\gamma}$ is the present photon temperature and $M_{Pl} = 1.22 \times 10^{19} \GeV$. The prefactor 3 accounts for the three colours of $\pb$. 
The $\pb$ mass is therefore related to the decay temperature and $r_{BDM}$ by
\be{e6b} m_{\pb} = 2.81 \TeV  \times g(T_{d})^{1/4} r_{BDM}^{1/2}(B(\pb) + B(\pbh))^{-1/2}  \left(\frac{T_{d}}{1 \GeV}\right)^{1/2} \lambda_{B} \left(1 - 
\frac{m_{S}^{2}}{m_{\pb}^{2}} \right)^{1/4}    ~.\ee

\subsection{Dark Matter Density}

   The annihilation cross-section times relative velocity for gauge singlet scalar dark matter \cite{gsdm3,gsdm4,gsdm5} is summarized in Appendix B. If $T_{d} < T_{s}$ then the density of dark matter is a non-thermal density from $\pb$, $\pbh$ annihilation. 
The total density of $s$ and $\sp$ dark matter is then given by  
\be{e15}  \Omega_{DM} =  \frac{2 m_{s}}{\rho_{c}} \frac{g(T_{\gamma})}{g(T_{d})^{1/2}} \left( \frac{4 \pi^3}{45 M_{Pl}^2} \right)^{1/2} \frac{T_{\gamma}^{3}}{T_{d}} \frac{1}{\left< \sigma v \right>_{s}}   ~.\ee
If $T_{d} > T_{s}$ then the dark matter is purely thermal relic in nature. In this case we replace $T_{d}$ by $T_{s}$ in \eq{e15}. 

\section{Results}

    Our aim is to understand why the baryon and dark matter densities are within an order of magnitude of each other. We therefore compute $r_{BDM}$ as a function of the inputs $m_{s}$, $m_{\pb}$, $\lambda_{s}$, $\lambda_{B}$ and $T_{d}$ and study how large a region of the parameter space can account for values of $r_{BDM}$ in the range 0.1 to 10 when $\Omega_{s} = 0.23$. 
The main constraints on the model are that (i) $m_{\pb} > m_{s}$, so that $\pb \pbh$ annihilation to $s \sp$ is kinematically allowed, and (ii) that $T_{d} < T_{\pb}$, so that the B-violating interaction is out of thermal equilibrium and cannot erase the asymmetry in $\pb$ and $\pbh$. We set the Higgs mass to $m_{h} = 150 \GeV$ and $ B(\pb) + B(\pbh) = 1$ throughout.

     We first consider the constraint on scalar masses when the couplings are fixed to have values $\lambda_{s} = \lambda_{B} = 0.1$.  In Figure 1 we show $m_{\pb}$ for the cases $r_{BDM} = 0.1$, 1.0 and 10.0 and $m_{s}$ leading to $\Omega_{DM} = 0.23$ when $T_{d} = 0.1 - 10 \GeV$. We also show the observed baryon-to-dark matter ratio, $r_{BDM} = 0.2$. In Figure 2 we show the same for $T_{d} = 1 - 80 \GeV$.  
A wide range $m_{\pb}$ is seen to be compatible with $r_{BDM}$ being within an order of magnitude of unity, from O(1) TeV to a few tens of TeV for $T_{d} \approx 50 \GeV$ and from  O(100) GeV to a few TeV for $T_{d} \lae 1$ GeV. Interestingly, the observed value of $r_{BDM}$ favours lower values of $m_{\pb}$, less than 2 TeV for $T_{d} \lae 80 \GeV$ and less than 1 TeV for $T_{d} \lae 10 \GeV$, improving the prospects for production of $\pb$ and $\pbh$ at the LHC.  

    We find that in general $m_{s} < m_{\phi}$ when $\Omega_{DM} = 0.23$. There are multiple solutions for $m_{s}$ with $\Omega_{DM} = 0.23$ for a given $T_{d}$. This is more clearly seen in Figure 3, which shows $m_{s}$ as a function of $T_{d}$. The large $m_{s}$ branch 
is primarily due to annihilation to $WW$ and $ZZ$. In general, the freeze-out temperature is given by $T_{s} \approx m_{s}/25$, with a similar result for $T_{\pb}$. 
Therefore the upper branch has $T_{d} < T_{s}$ and so the $s$ dark matter density in this case is non-thermal. There are also two lower branches; one slightly larger than the Higgs pole at $m_{s} \approx 79 \GeV$ when $T_{d} \lae 2 \GeV$, and a second at $m_{s} \approx 67 \GeV$. For $T_{d} \gae 3 \GeV$ we find that $T_{s} < T_{d}$ in this case, in which case the lower branch $s$ density is a thermal relic density determined by annihilation to primarily b quark pairs. Both of the lower branches require that $m_{s}$ is close to the Higgs pole $m_{h}/2$, so these solutions appear less likely than the more generic heavy $m_{s}$ solution, in which case the dark matter is most likely to be non-thermal from $\pb \pbh$ annihilation.

  We next consider the case where the masses are fixed to show the effect of varying $\lambda_{B}$ and $\lambda_{s}$. In Figure 4 we show the case with $m_{s} = 120 \GeV$ and $m_{\pb} = 400 \GeV$ when $T_{d} < 10 \GeV$. In Figure 5 we show the same for the case $ 1 \GeV < T_{d} < 80 \GeV$. (This range of $m_{s}$ may be observable in the near future at direct dark matter detection experiments, while $m_{\pb}$ should be accessible to the LHC.)  
For these masses, values of $\lambda_{B}$ in the range 0.004 to 0.06 will produce $r_{BDM}$ in the range 10 to 0.1 when $T_{d} \approx 10 \GeV$. The range of $\lambda_{B}$ is 0.01 to 0.6 when $T_{d} \lae 1 \GeV$. For $T_{d} \approx 50 \GeV$ the range of $\lambda_{B}$ is 0.003 to 0.03. Smaller $r_{BDM}$ favours larger $\lambda_{B}$. 
The dark matter density $\Omega_{DM} = 0.23$ requires $\lambda_{s} \approx 0.04$ once $T_{d} \gae 5 \GeV$. The plot of $\lambda_{s}$ is $T_{d}$ independent once $T_{d} \gae 5 \GeV$, since in this case the dark matter is produced thermally.  For $T_{d} \lae 1 \GeV$, $\lambda_{s} \gae 0.1$.

     In summary, if the baryon asymmetry is injected at $0.1 \GeV \lae T_{d} \lae 100 \GeV$ (not a very narrow range), then for masses characterized by the weak to TeV scale (a natural range for SM extensions) and couplings in the range 0.001-1 (not unusually small), $r_{BDM}$ is within an order of magnitude of unity. The observed $r_{BDM}$ favours larger couplings and smaller $m_{\pb}$. For $\lambda_{s} \approx \lambda_{B} \approx 0.1$ and $T_{d} \lae 10 \GeV$ it is quite natural to have $m_{\pb} \lae 500 \GeV$ and $m_{s} \lae 200 \GeV$ when $r_{BDM} = 0.2$, in which case production of annihilons at the LHC and direct detection of $s$ dark matter may be possible.

\begin{figure}[htbp]
\begin{center}
\epsfig{file=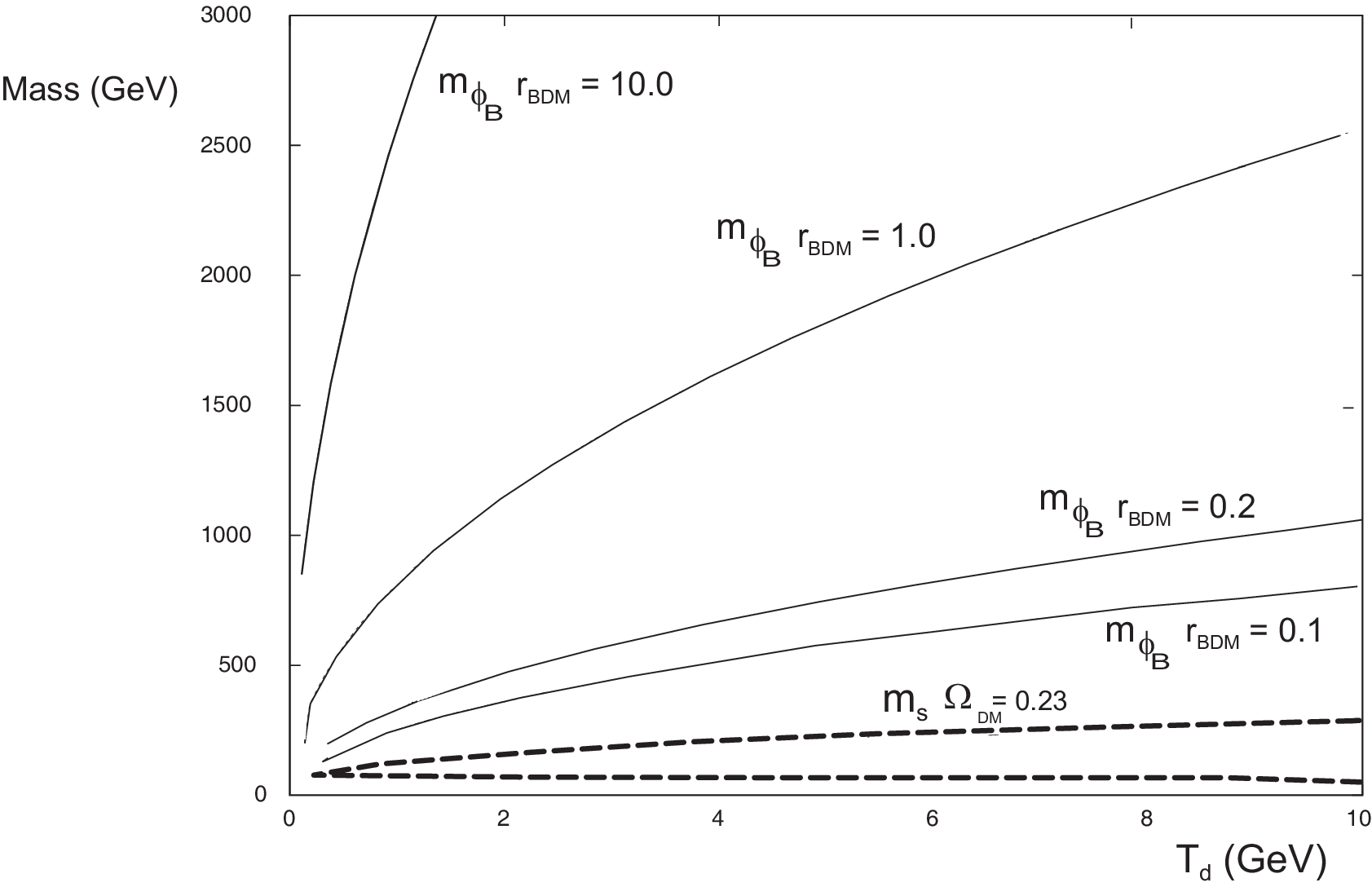, width=0.5\textwidth, angle = -90}
\caption{Values of $m_{\pb}$ for different $r_{BDM}$ (solid lines) and $m_{s}$ for $\Omega_{DM} = 0.23$ (dashed line) as a function of $T_{d}$ for the case $\lambda_{B} = \lambda_{s} = 0.1$.}
\label{fig1a}
\end{center}
\end{figure}

\begin{figure}[htbp]
\begin{center}
\epsfig{file=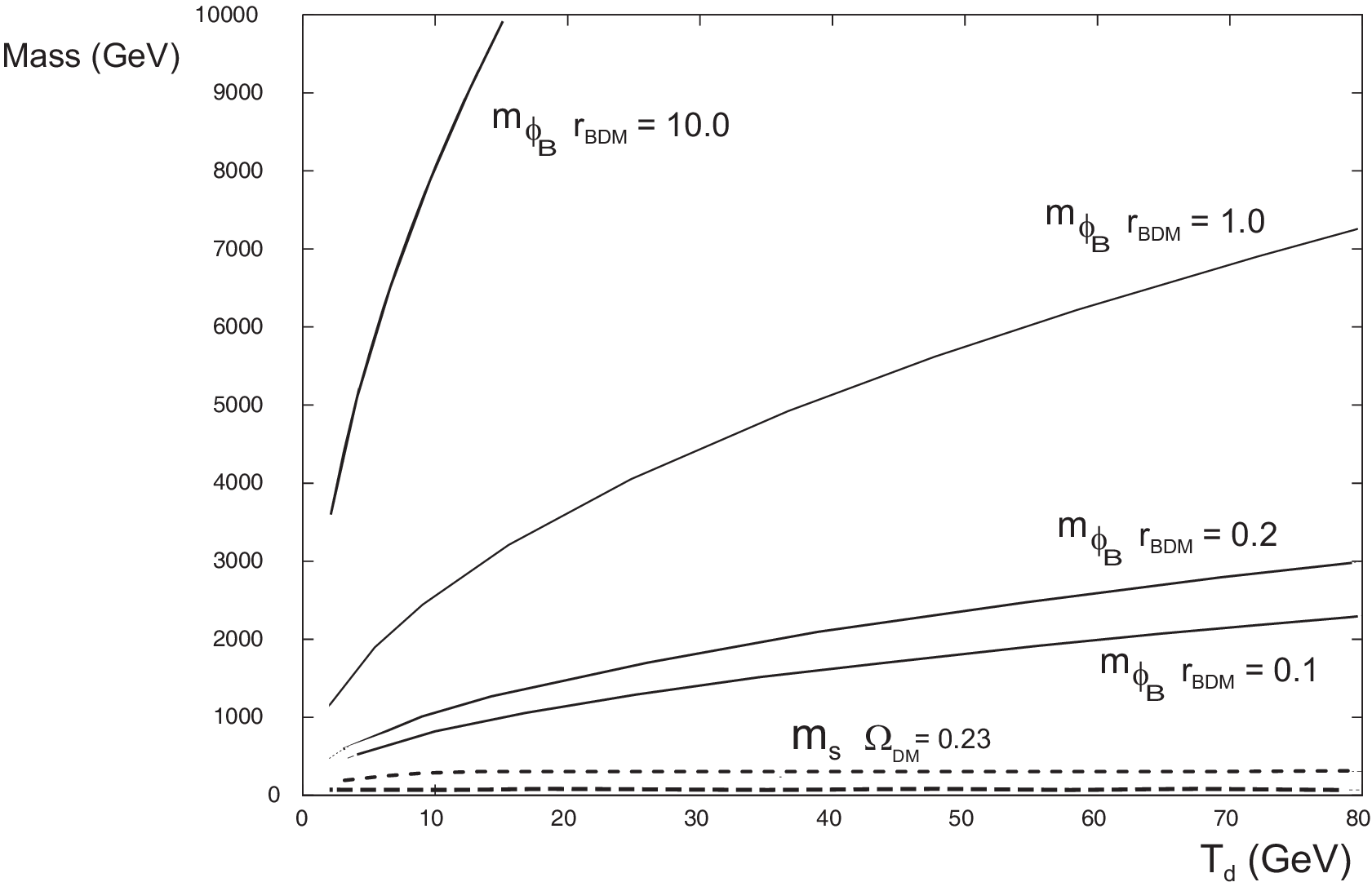, width=0.5\textwidth, angle = -90}
\caption{Values of $m_{\pb}$ for different $r_{BDM}$ (solid lines) and $m_{s}$ for $\Omega_{DM} = 0.23$ (dashed line) as a function of $T_{d}$ for the case $\lambda_{B} = \lambda_{s} = 0.1$.}
\label{figx1}
\end{center}
\end{figure}

\begin{figure}[htbp]
\begin{center}
\epsfig{file=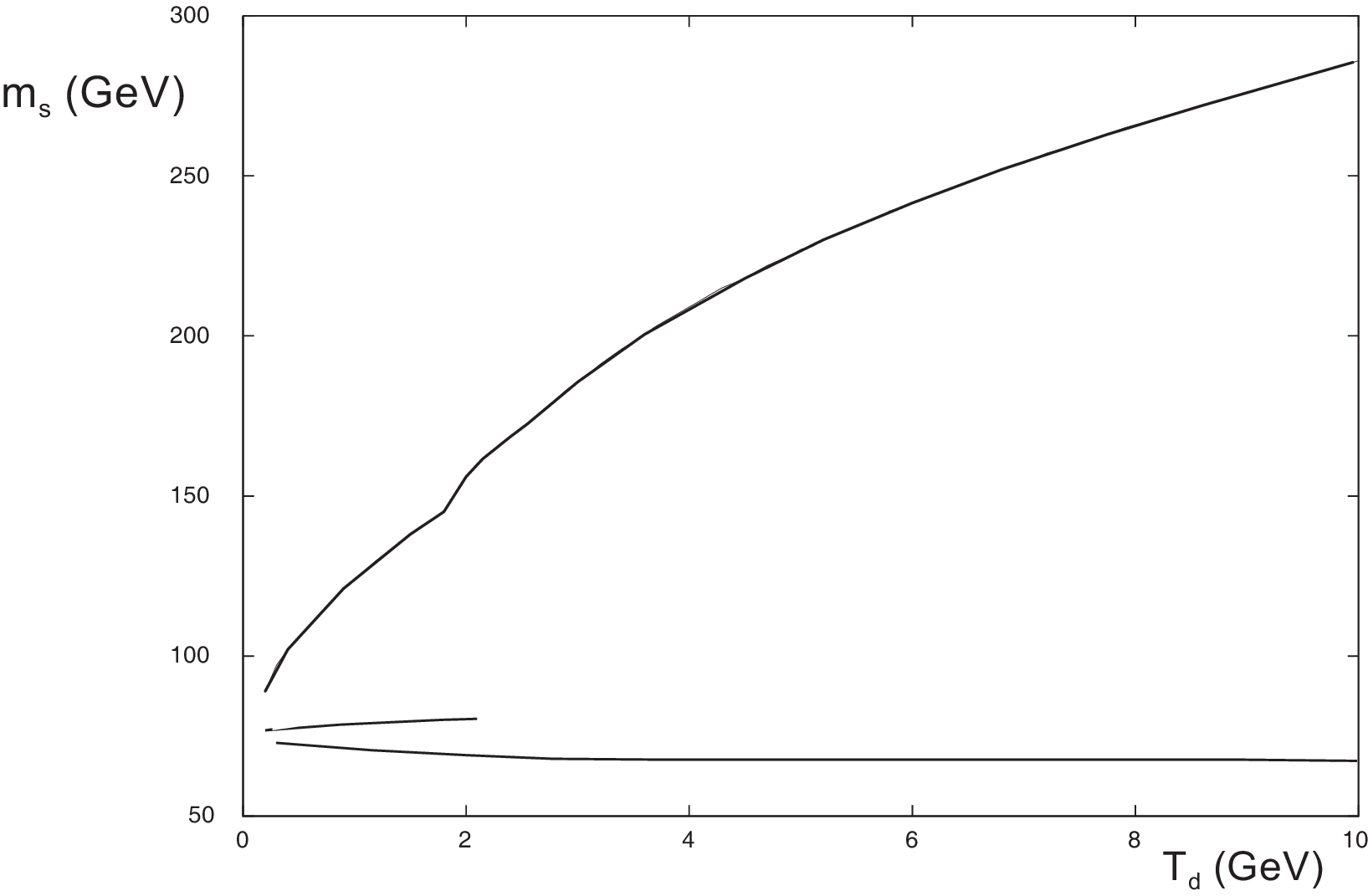, width=0.5\textwidth, angle = -90}
\caption{Values of $m_{s}$ for $\Omega_{DM} = 0.23$ as a function of $T_{d}$ for the case $\lambda_{B} = \lambda_{s} = 0.1$.}
\label{fig1b}
\end{center}
\end{figure}

\begin{figure}[htbp]
\begin{center}
\epsfig{file=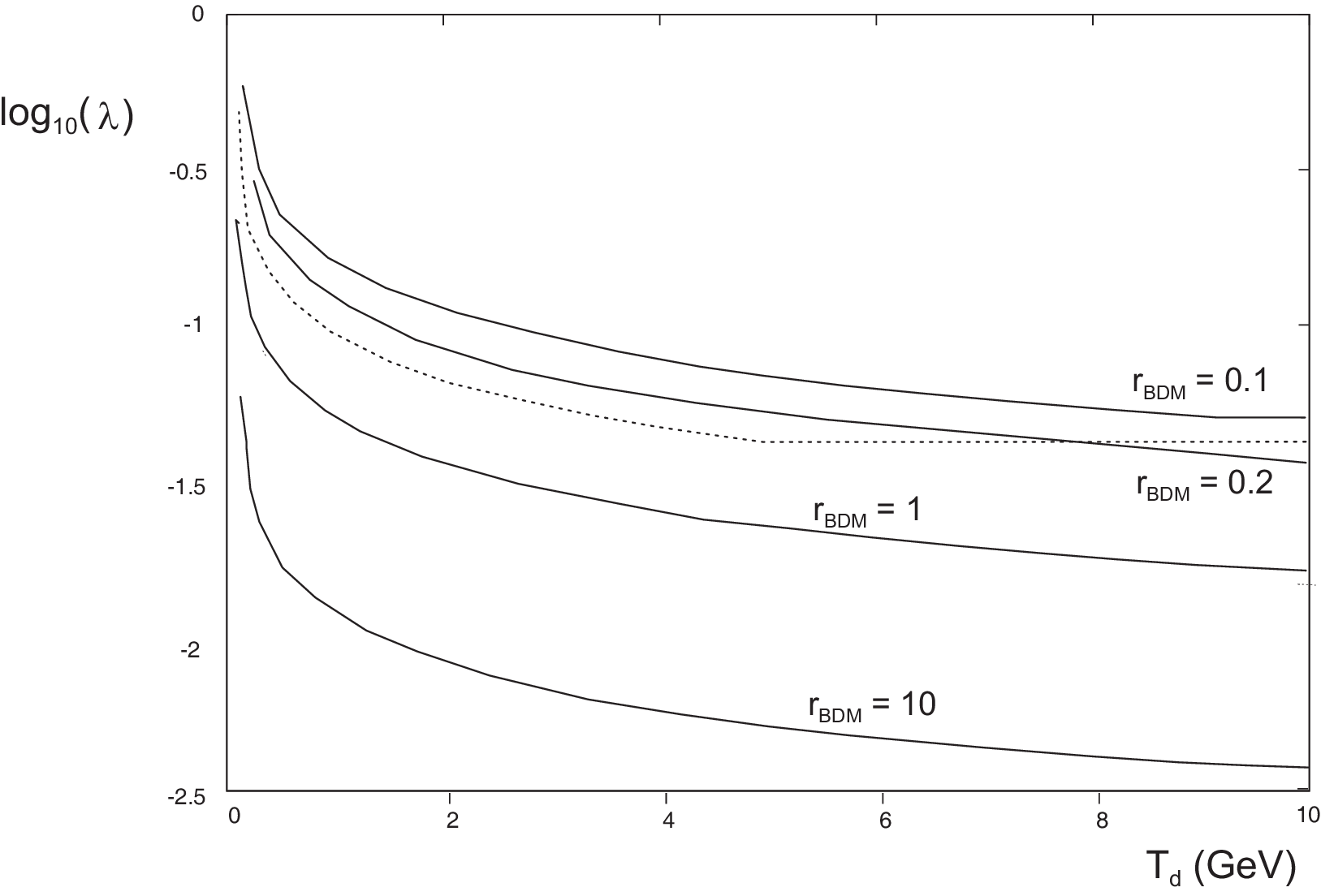, width=0.5\textwidth, angle = -90}
\caption{Values of $\lambda_{B}$ for different $r_{BDM}$ (solid lines) and $\lambda_{s}$ for $\Omega_{DM} = 0.23$ (dotted line) as a function of $T_{d}$ for the case $m_{\pb} = 400 \GeV$ and $m_{s} = 120 \GeV$.}
\label{fig4c}
\end{center}
\end{figure}

\begin{figure}[htbp]
\begin{center}
\epsfig{file=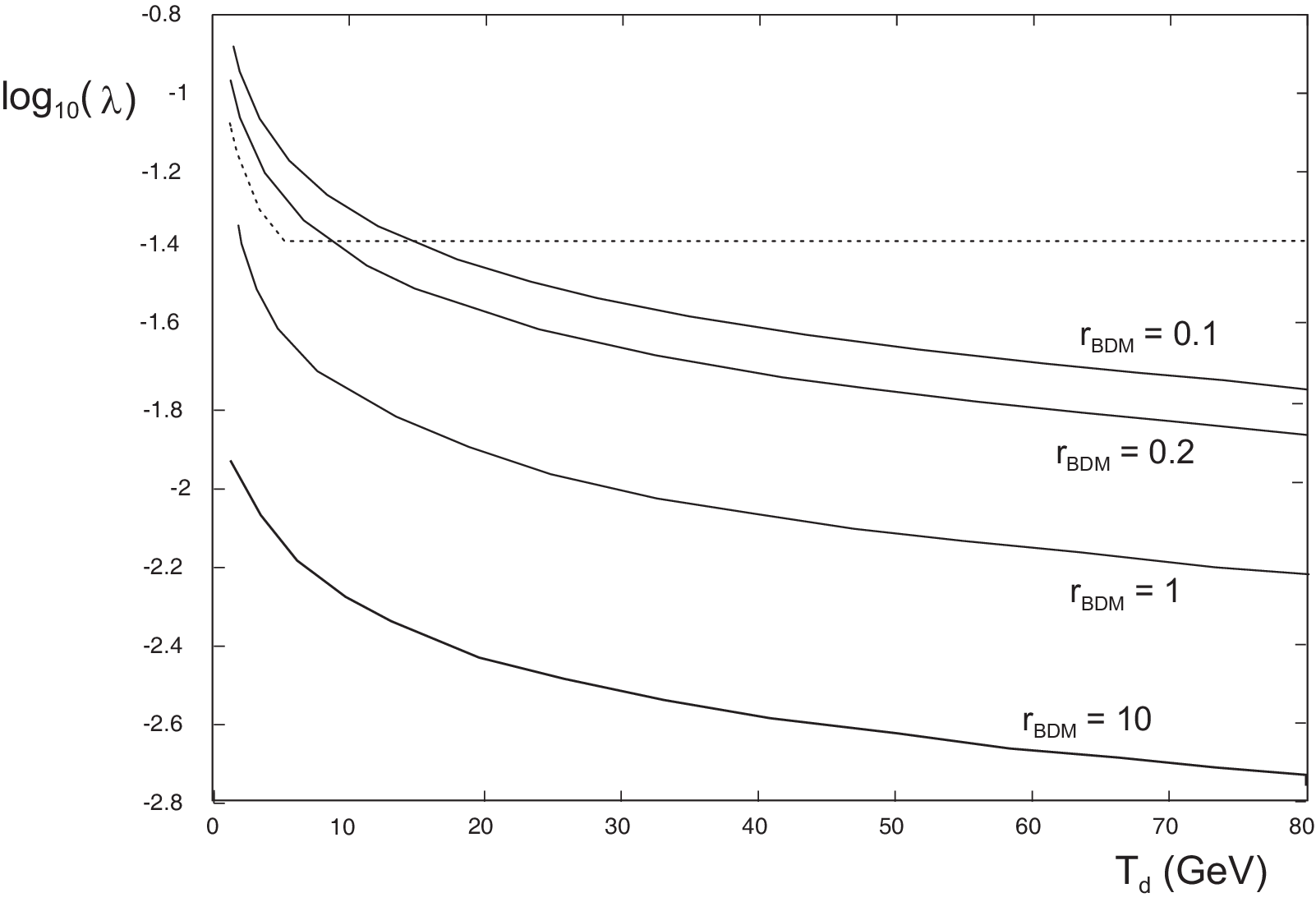, width=0.5\textwidth, angle = -90}
\caption{Values of $\lambda_{B}$ for different $r_{BDM}$ (solid lines) and $\lambda_{s}$ for $\Omega_{DM} = 0.23$ (dotted line) as a function of $T_{d}$ for the case $m_{\pb} = 400 \GeV$ and $m_{s} = 120 \GeV$.}
\label{fig4d}
\end{center}
\end{figure}

\section{Annihilon decay} 

              In the previous section we showed that the baryon asymmetry and dark matter density from $\pb \pbh$ annihilation can be naturally similar to each other and to a thermal relic WIMP density. However, we still need to transfer the $\pb \pbh$ asymmetry to a conventional baryon asymmetry. At this stage the $\pb \pbh$ asymmetry does not necessarily correspond to a baryon asymmetry. This will be determined by the decay modes of $\pb$ and $\pbh$ to quarks. There are two possibilities: (i) baryon number is conserved by the model as a whole and 
$\pb$ and $\pbh$ have specific baryon numbers, or (ii) baryon number is conserved only by the SM sector (to ensure proton stability) and $\pb$ and $\pbh$ can decay to final states with different baryon numbers. In this case the effective baryon number of $\pb$ and $\pbh$ will be determined by their dominant decay mode to quarks. Since (i) is essentially a subset of (ii), we will concentrate on the second possibility.

                The lifetimes of $\pb$ and $\pbh$ are necessarily long. Defining their decay temperature to SM quarks and leptons to be $T_{D}$, we require that $1 \MeV \lae T_{D} \lae T_{d}$, where the lower bound is from nucleosynthesis and the upper bound from the requirement that the initially large $\pb \pbh$ asymmetry annihilates down to a WIMP-like density at $T_{d}$ prior to decaying to quarks. Therefore \cite{bm1}  
\be{e16}    1.5\; {\rm s} \gae \tau \gae 8 \times 10^{-11} \left(
\frac{100 \GeV}{T_{d}} \right)^2  \;\; s  ~.\ee
The long annihilon life-time requires either an extremely small renormalizable Yukawa coupling of the form $\lambda \pb \overline{\psi} \psi$ to SM fermions $\psi$, 
\be{e17} \lambda \lae 1.2 \times 10^{-10} \left( \frac{T_{d}}{1 \GeV}\right)
 \left( \frac{1 \TeV}{m_{\pb}}\right)^{1/2}   ~,\ee
  or a non-renormalizable coupling suppressed by a sufficiently large mass scale. The former possibility appears explicitly unnatural, so we will consider the latter. In this case we need to explain why there are no renormalizable couplings leading to rapid $\pb$ decay.  

    The simplest way to achieve this is to assume that the annihilons have a large hypercharge. The largest hypercharge carried by a pair of SM fermions has magnitude $|Y| = 2$ (for $\overline{e_{R}^{c}}e_{R}$), while the largest combinations carrying baryon number are $\overline{u_{R}}u_{R}^{c}$ and 
$\overline{d_{R}}e_{R}^{c}$, with $|Y| = 4/3$. 
The SM fermion pairs which transform as $({\bf 3}, {\bf 1})$ or $({\bf \overline{3}}, {\bf 1})$ have hypercharge $|Y| = 1/3, 2/3$ or 4/3. Therefore if $Y(\pb) = 5/3$ in the case where $\pb$ transforms as $({\bf 3, 1})$,  there  are no renormalizable couplings of $\pb$ to SM fermions\footnote{Note that inclusion of the Higgs doublet can only increase the dimension of an operator relative to the case without the Higgs, since the Higgs must occur in isosinglet combinations such as $H^{\dagger}Q$ and $HQ$, which have hypercharge equal in magnitude to SM fermions and so can be replaced by SM fermions.}. However, non-renormalizable couplings of $\pb$ and $\pbh$ to $d=6$ operators are possible, for example\footnote{The colour indices of the three triplets are contracted by the anti-symmetric tensor to form a $SU(3)_{c}$ singlet.} \be{e18} \frac{1}{M^3}\pb \overline{d_{R}^{c}} d_{R} \overline{L_{L}^{c}} L_{L}   ~\ee
and 
\be{e19}  \frac{1}{M^3}\pbh \overline{d_{R}} e_{R}^{c} Q_{L} Q_{L}  ~.\ee 
The mass $M$ should then be in the range $10^{6}-10^{8} \GeV$ to account for the low decay temperature $T_{D}$ \cite{bm1}. Note that for $\pbh$ to decay, we must assume that $Z_{A}$ is slightly broken by the non-renormalizable operators. However, since these operators are suppressed by a large mass scale, this small breaking of $Z_{A}$ will not introduce any dangerous mass mixing between $\pb$ and $\pbh$.
If the operators \eq{e18} and \eq{e19} are dominant, the effective baryon number of $\pb$ and $\pbh$ would be $B(\pb) = -2/3$ and $B(\pbh) = -1/3$.  
However, if we do not assume baryon number conservation, there are other possible operators, for example 
\be{e20} \frac{1}{M^3}\pb \left( \overline{e_{R}} Q_{L} \overline{e_{R}} L_{L} \right)^{\dagger}  ~,\ee
which allows $\pb$ to decay to a final state with $B = 1/3$. In this case the effective baryon number of $\pb$ and $\pbh$ will be determined by their dominant decay modes. 
Production of long-lived scalars with large hypercharge at the LHC, decaying to baryon number and possibly with baryon number violation in their decay 
modes, would therefore support this class of model.

\section{Conclusions}

      The similarity of the observed values of $\Omega_{B}$ and $\Omega_{DM}$ 
points to two distinct coincidence problems: why the density of baryons and dark matter are similar to each other and why they are both similar to a typical thermal relic WIMP density (the 'WIMP miracle'). We have shown that it is possible to explain these coincidences via a simple extension of the SM based on gauge singlet scalar dark matter and colour triplet scalar annihilons. A $Z_{2}$ discrete symmetry ($Z_{A}$) and new scalar particles are necessary to prevent dangerous B-violating interactions. The new scalars then provide a natural dark matter candidate if stabilized by a second $Z_{2}$ symmetry, $Z_{S}$. 
The model predicts a pair of stable scalar dark matter particles in the case where the scalars are equal in mass. The mechanism determining the final baryon asymmetry ('baryomorphosis') is based on the injection of a large baryon asymmetry in scalar annihilons at a relatively low temperature ($0.1 \GeV \lae T_{d} \lae 100 \GeV$), which subsequently annihilate via a B-violating interaction to a thermal relic WIMP-like density of baryons.  
For couplings $\lambda_{B} \sim 0.01-1$ and annihilon masses $m_{\phi} \sim 100 \GeV - 10 \TeV$, which are the ranges we might expect for a TeV-scale extension of the SM, 
the value of $\Omega_{B}/\Omega_{DM}$ is naturally within an order of magnitude of unity. 
Therefore the initial large baryon asymmetry is converted to both a thermal WIMP-like baryon asymmetry and a thermal WIMP-like scalar dark matter density. Both densities are typically non-thermal, but both are determined by broadly weak strength annihilation processes and so are naturally similar to a thermal relic WIMP density. The observed baryon to dark matter ratio favours lower masses and larger couplings for the annihilons, favouring production at the LHC, and lower dark matter singlet masses, which might be observed in direct dark matter detection experiments. 

    The asymmetry in the annihilons is transferred to a conventional baryon asymmetry by decay to SM fermions. This must occur at a low temperature, implying a long lifetime. This suggests that renormalizable couplings of the annihilons to SM fermions must be highly suppressed or eliminated. This is most easily achieved by assigning a large hypercharge ($|Y| > 4/3$) to the annihilons. The annihilon asymmetry does not necessarily directly correspond to a baryon asymmetry. One possibility is that baryon number is conserved only in the SM sector. In this case annihilons could decay to final states with different baryon number, with the effective baryon number of the annihilons being determined by their dominant decay mode. Observation of pairs of long-lived scalars with mass O(100) GeV to a few TeV, with opposite gauge charge but possibly different mass, and with large hypercharge and possibly B-violating decay modes, would therefore strongly support the class of model we have presented here.        

    It is important to consider whether such models are plausibly natural extensions of the SM. The model we have presented is a simple scalar particle extension of the SM with a $Z_{2} \times Z_{2}$ discrete symmetry, which requires no fine-tuning of masses and couplings. It should be emphasized that the SM already requires an additional dark matter particle stabilized by a symmetry if dark matter is due to a WIMP, so the model might be considered an extension of this concept. The key requirements of the model are a relatively low temperature for the injection of the annihilon asymmetry and for its subsequent decay to a conventional baryon asymmetry. However, a significantly wide range of injection temperature ($ 0.1 \GeV \lae T_{d} \lae 100 \GeV$) is compatible with $\Omega_{B}$ being within an order of magnitude of $\Omega_{DM}$ for natural scalar masses and couplings. Therefore although there is a requirement for a non-trivial sequence of processes to take place, there is nothing overtly unnatural in the requirements of the model.

   The question of whether there exists a natural mechanism to relate the density of baryons and dark matter to a thermal relic WIMP density is fundamentally important to our understanding of the origin of baryons and dark matter. The model we have presented demonstrates that it is not necessary to invoke anthropic selection to explain the baryon asymmetry when dark matter is explained by the WIMP miracle. Since the new physics required is broadly at the weak or TeV scale, we can hope that experiment  will be able to clarify the nature of the observed coincidence of the baryon and dark matter densities.

\renewcommand{\theequation}{A-\arabic{equation}}
 \setcounter{equation}{0} 

\section*{Acknowledgement} The author would like to thank Rose Lerner for comments. 

\section*{Appendix A: Scattering rates from the thermal background and the slowing of relativistic scalars}

     In our discussion of the relic density we have assumed that particles can rapidly become non-relativistic before annihilating. Here we show that this is the case. For the case of charged annihilons, we consider the scattering rate of the annihilons from thermal background photons and show that this is rapid compared with the expansion rate and the   
$\phi_{B} \pbh$ annihilation rate. For the case of the relativistic gauge singlet $s$ and $\sp$ particles produced by $\phi_{B} \pbh$ annihilation, we show that as long as relativistic c-quarks are in thermal equilibrium, scattering with SM fermions mediated by Higgs exchange can slow the singlet scalars sufficiently rapidly compared with the annihilation and expansion rate to justify treating their annihilations as non-relativistic.

\subsection{Scattering and slowing of charged relativistic annihilons} 

   For the case of charged annihilons, we consider the scattering from photons in the thermal background. The interaction term is 
\be{a1}  {\cal L}_{int} = e^2 Q^2 A_{\mu} A^{\mu} \pb^{\dagger} \pb   ~,\ee
where $Q$ is the electric charge of the annihilon.
The scattering cross-section is then
\be{a2} \sigma = \frac{e^4 Q^4}{2 \pi s}    ~.\ee 
In the thermal rest frame, we consider the energy of the photons on average to be $E_{T} \approx 3T$ and we define the energy of the $\phi_{B}$, $\pbh$ to be $E$, where $E$ is assumed large enough that the $\pb$ and $\pbh$ are relativistic. In this case $s \approx  4 E E_{T} + m_{\pb}^2$.  
The condition for the scattering process to efficiently slow the $\pb$ particles is that 
\be{an1}  \frac{\Gamma_{sc}}{H} \frac{\Delta E}{E} \gae 1   ~,\ee
where $\Delta E$ is the energy loss per scattering and $\Gamma_{sc} = n \sigma$ is the scattering rate of the relativistic $\pb$ particles from photons, where 
$n \approx 2T^3/\pi^2$ is the  thermal photon number density. 
If this is satisfied then the $\pb$ will lose most of their energy in a time shorter than $H^{-1}$. $\Delta E/E$ will depend on whether the $\pb$ particles are relativistic 
in the CM frame, which is true if $4E E_{T} > m_{\pb}^2$. In this case $s = 4 E E_{T}$, the average energy transfer per scattering is $\Delta E= E/2$  and the 
condition for efficient loss of energy becomes
\be{an2} E \lae \frac{e^4 Q^4 M_{Pl}}{24 \pi^3 k_{T}} \approx 1 \times 10^{13} \; Q^4 \GeV    ~,\ee
where $k_{T} = (4 \pi^3 g(T)/45)^{1/2}$ and we use $g(T) \approx 100$.  
This is easily satisfied so long as the initial energy of the $\pb$ is not very large.
Therefore the $\pb$ will lose energy until they become non-relativistic in the CM frame. Once non-relativistic in the CM frame, $s = m_{\pb}^2$ and $\Delta E/E = 2E E_{T}/m_{\pb}^2$. 
The condition for efficient loss of energy then becomes 
\be{an3} \frac{6 e^4 Q^4 M_{Pl} E T_{d}^2}{\pi^3 k_{T} m_{\pb}^4} \gae 1   ~.\ee
This is most difficult to satisfy when $E \rightarrow m_{\pb}$, in which case the condition becomes 
\be{an4} T_{d} \gae 8 \times 10^{-4} \frac{1}{Q^2} \left( \frac{m_{\pb}}{1 \TeV}\right)^{3/2}  \GeV ~.\ee
This is satisfied for $T_{d} \gae 1 \MeV$. Therefore, for the range of $T_{d}$ of interest to us here, the initially relativistic $\pb$ will become non-relativistic on a timescale 
short compared with $H^{-1}$.  Since the freeze-out 
of the non-relativistic $\pb \pbh$ annihilation cross-section occurs once $\Gamma_{ann} \approx H$, the annihilons 
will become non-relativistic before they freeze-out.

\subsection{Scattering and slowing of gauge singlet scalars}

   In the case of gauge singlets, the interaction with the thermal background can be much weaker, in particular at low $T_{d}$ when only light fermions with small Yukawa couplings are a significant component of the thermal background. 

    For gauge singlets interacting with the SM via the interaction
$(\lambda_{s}/2) s^2 H^{\dagger} H$, once $<h^{o}> = 246 \GeV$ is introduced 
there is a t-channel Higgs exchange interaction with SM fermions. The average squared matrix element computed in the CM frame is 
\be{a8} \overline{|{\cal M}|}^2  = \frac{ 2 \lambda_{s}^2 \lambda_{f}^2 <h^{o}>^2 k^2 (1 - \cos \theta) }{\left( 2 k^2 \left(1 - \cos \theta \right) + m_{h}^2 \right)^2 }    ~,\ee 
where $\lambda_{f}= m_{f}/<h^{o}>$ is the Yukawa coupling of SM fermion $f$, $\theta$ is the scattering angle in the CM frame and $k$ is the $s$ momentum in the CM frame, given by the solution of $2 m_{s} k + k^2 = 4 E E_{T}$.  
The cross-section is then
\be{a10}   \sigma = 
\frac{\lambda_{s}^2 \lambda_{f}^2 <h^{o}>^2 \alpha_{sc}}{64 \pi s k^2} 
  ~,\ee  
where 
\be{a10a} \alpha_{sc} = \ln \left( 1 + \frac{4 k^2}{m_{h}^2} \right) + 
\left( 1 + \frac{4 k^2}{m_{h}^2} \right)^{-1}  - 1   ~.\ee
If $4 k^2/m_{h}^2 \gg 1$ then 
\be{a11} \alpha_{sc} \approx \ln \left(\frac{4 k^2}{m_{h}^2} \right) - 1   ~,\ee
while if $4 k^2/m_{h}^2 \ll 1$ then
\be{a12} \alpha_{sc} \approx \frac{8 k^4}{m_{h}^4}  ~.\ee 
We consider the limit where $T_{d}$ is low compared with $m_{s}$ and $m_{h}$ and the $s$ energy $E \rightarrow m_{s}$, which will give the least efficient 
transfer of energy.  (We have checked that no stronger constraint results from considering $E > m_{s}$.) In this limit, we expect $E < m_{s}^2/4 E_{T}$ and so the $s$ will be non-relativistic in the CM frame, in which case $\Delta E/E = 2 E E_{T}/m_{s}^2$ and $k = 2 E E_{T}/m_{s} \ll m_{h}/2$. Therefore the condition for efficient loss of energy becomes
\be{a13} \frac{\lambda_{s}^2 \lambda_{f}^2 <h^{o}>^2 k^3 T_{d} M_{Pl}}{2 \pi^3 m_{h}^4 m_{s}^3 k_{T}} \gae 1 ~.\ee
With $k = 6 T_{d} E/m_{s}^{2}$ and $E = m_{s}$, this becomes
\be{a14} T_{d} \gae 0.33 \GeV \times \left(\frac{5 \times 10^{-3}}{\lambda_{f}}\right)^{1/2}  \left(\frac{0.1}{\lambda_{s}}\right)^{1/2} 
\left(\frac{m_{h}}{150 \GeV}\right) \left(\frac{m_{s}}{100 \GeV}\right)^{3/4}   ~,\ee
where we have normalized $\lambda_{f}$ to the c-quark Yukawa coupling $\lambda_{c} = 5 \times 10^{-3}$. c-quarks will form part of the relativistic thermal bath if 
$T_{d} \gae m_{c}/3 = 0.4 \GeV$, therefore since in this case the bound from \eq{a14} is $T_{d} \gae 0.3 \GeV$, the energy loss will be efficient and so $s$ particles will become non-relativistic on a timescale short compared with $H^{-1}$. In this case the $s$ scalars will efficiently lose energy to the thermal background and become non-relativistic before they annihilate.
 However, if $T_{d} < 0.4 \GeV$ then scattering must be through s-quarks and muons. In this case $\lambda_{f} \approx 4 \times 10^{-4}$ and so the bound from \eq{a14} becomes $T_{d} \gae 1.2 \GeV$. Therefore energy loss through scattering will be 
ineffective and so the $s$ particles will annihilate while relativistic. In this case the $s$ particles become non-relativistic only via redshifting of their momentum. The final $s$ density will therefore be fixed at the temperature $T_{NR}$ at which they become non-relativistic rather than at $T_{d}$, resulting in an enhancement of the relic $s$ density by a factor $T_{d}/T_{NR}$. This would require modification of our results for $m_{s}$ and $\lambda_{s}$  at very low $T_{d}$, with $m_{s}$ suppressed by a factor $T_{NR}/T_{d}$ for a given $\Omega_{DM}$. 

\renewcommand{\theequation}{B-\arabic{equation}}
 \setcounter{equation}{0} 

\section*{Appendix B: Gauge Singlet Scalar Annihilation Cross-Section}

   For convenience we provide the non-relativistic annihilation cross-section times relative velocity for gauge singlet scalars which we used in the calculation of the gauge singlet relic density. This has been discussed in \cite{gsdm3,gsdm4,gsdm5}.  
The tree-level processes contributing to $ss$ annihilation are (i) $ss \rightarrow hh$, (ii) $ss \rightarrow WW$, (iii) $ss \rightarrow ZZ$ and (iv) $ss \rightarrow \overline{f}f$, where $f$ is a Standard Model fermion.  (The cross-sections for $\sp \sp$ annihilation are similar.) (i) proceeds via a 4-point contact interaction, an s-channel Higgs exchange interaction and a t- and u-channel $s$ exchange interaction. The resulting $\langle\sigma v_{rel}\rangle$ is
\bea \langle\sigma v_{rel}\rangle_{hh}& =& \frac{\lhs^{2}}{64 \pi m_s^{2}}
\left[ 1 + \frac{3 m_{h}^2}{\left(4 m_s^2 - m_h^2\right) } + \frac{2 \lhs v^2}{\left(m_h^2 - 2 m_s^2\right) }\right]^2 \nonumber \\
& & \times \left(1-\frac{m_{h}^{2}}{m_s^{2}}\right)^{1/2}~.
\eea
$SS \rightarrow WW,\;ZZ,\;\overline{f}f$ all proceed via s-channel Higgs exchange. The corresponding
$\langle\sigma v_{rel}\rangle$ are:
\bea
\langle\sigma v_{rel}\rangle_{WW}& =& 2\left(1+\frac{1}{2}\left(1-\frac{2 m_s^{2}}{m_{W}^{2}}\right)^{2} \right)\left(1-\frac{m_{W}^{2}}{m_s^{2}}\right)^{1/2} \nonumber \\
& & \times \frac{\lhs^{2} m_{W}^{4}}{8 \pi m_s^{2}\left(\left(4 m_s^{2} -m_{h}^{2}\right)^{2}+m_{h}^{2}\Gamma_{h}^{2}\right)  }
~,\eea
\bea
\langle\sigma v_{rel}\rangle_{ZZ}& =&  2\left(1+\frac{1}{2} \left(1-\frac{2 m_s^{2}}{m_{Z}^{2}} \right)^{2}\right)\left(1-\frac{m_{Z}^{2}}{m_s^{2}}\right)^{1/2} \nonumber \\
& & \times \frac{\lhs^{2} m_{Z}^{4}}{16 \pi m_s^{2} \left(\left(4 m_s^{2} -m_{h}^{2}\right)^{2}+m_{h}^{2}\Gamma_{h}^{2}\right)  }
~\eea
and
\be{b1}  \langle \sigma v_{rel}\rangle_{ff} =\frac{m_{W}^{2}}{\pi g^{2}} \frac{\lambda_{f}^{2}
\lhs^{2} }{\left(\left(4m_s^{2}-m_{h}^{2}\right)^{2}+m_{h}^{2}\Gamma_{h}^{2}\right) }
\left(1-\frac{m_{f}^{2}}{m_s^{2}}\right)^{3/2}
~.\ee
Here the fermion Yukawa coupling is $\lambda_{f} =
m_{f}/v$, where $v = 246$~GeV and $m_{f}$ is the fermion
mass (fermions should be summed over colours).
 $\Gamma_{h}$ is the Higgs decay width.



\end{document}